\documentstyle[12pt]{article}

\newcommand{\pat}{\partial}
\newcommand{\be}{\begin{equation}}
\newcommand{\ee}{\end{equation}}
\newcommand{\bea}{\begin{eqnarray}}
\newcommand{\eea}{\end{eqnarray}}

\newcommand{\Fcal}{{\cal F}}
\newcommand{\Ucal}{{\cal U}}
\newcommand{\Vcal}{{\cal V}}
\newcommand{\hsf}{{\sf h}}
\newcommand{\half}{\frac{1}{2}}
\newcommand{\Xbar}{\bar{X}}
\newcommand{\xibar}{\bar{\xi }}
\newcommand{\barh}{\bar{h}}

\newcommand{\barone}{\bar{1}}
\newcommand{\bartwo}{\bar{2}}
\newcommand{\nbyn}{N \times N}

\newcommand{\Tr}{{\rm Tr}}
\newcommand{\tr}{{\rm tr}}

\newcommand{\unitm}{{\bf 1}}
\newcommand{\zerom}{{\bf 0}}

\newcommand{\bfE}{{\bf E}}

\newcommand{\bfM}{{\bf M}}

\begin{document}

\baselineskip 18pt

\begin{titlepage}
\begin{flushright}
CALT-68-2136 \\
September 1997
\end{flushright}

\vskip 1.2truecm

\begin{center}
{\Large {\bf Born-Infeld Actions 
   from Matrix Theory}}
\end{center}

\vskip 0.8cm

\begin{center}
{\bf Esko Keski-Vakkuri}$^1$ and {\bf Per Kraus}$^2$
\vskip 0.3cm
{\it California Institute of Technology \\
     Pasadena CA 91125, USA  \\
     email: esko or perkraus@theory.caltech.edu}
\end{center}

\vskip 2.2cm

\begin{center}
{\small {\bf Abstract: }}
\end{center}       
\noindent           
{\small 
We propose a formula for the effective action of Matrix Theory which
successfully reproduces a large class of Born-Infeld type D-brane probe
actions.  The formula is motivated by demanding consistency with known results,
and is tested by comparing with a wide range of source-probe calculations
in supergravity. In the case of D0-brane sources and Dp-brane probes, we study
the effect of boosts, rotations, and worldvolume electric fields on the probe,
and find agreement with supergravity to all orders in the gravitational 
coupling.  We also consider D4-brane sources at the one loop level and 
recover the correct probe actions for a D0-brane, and for a D4-brane rotated
at an angle with respect to the source.
}
\rm
\vskip 3.4cm

\small
\begin{flushleft}
$^1$ Work supported in part by a DOE grant DE-FG03-92-ER40701.\\    
$^2$ Work supported in part by a DOE grant DE-FG03-92-ER40701 and by the
DuBridge Foundation.
\end{flushleft}
\normalsize 
\end{titlepage}

\newpage
\baselineskip 16pt

\section{Introduction}

A remarkable feature of Matrix Theory (MT) \cite{BFSS} is 
its ability to describe a wide 
variety of objects within a single configuration space; the various known 
branes of string theory can be realized by choosing specific forms for the MT 
variables\footnote{There is, however, difficulty in realizing the transverse 
five brane, as Matrix theory appears to lack the needed central 
charge \cite{BSS}.}. 
One can also realize two widely separated branes, and study the resulting 
interactions by integrating out the massive degrees of freedom which couple 
the objects \cite{Bachas,DFS,KP,DKPS,Gil}. 
The results so obtained can be compared with the interactions 
predicted by supergravity, and can thus be used as checks of the MT proposal. 
During the past year, numerous checks of this sort have been performed, and 
have lent strong support to the accuracy of MT \cite{AB,LM,L2,L3,BC,BL,PP,DKM,
BFSS2,CT1,CT2,GR,Mal,KVK,P,BB,BBPT}. 
Typically, one computes one 
loop diagrams in Matrix quantum mechanics and compares the result against the 
leading long-distance interaction from supergravity. In the case of D0-brane 
scattering, one thereby verifies the coefficient of the well known $v^4/r^7$ 
term. In addition -- and even more impressively -- there have been 
calculations at the two loop level \cite{BB,BBPT} (see also \cite{GGR})
which have reproduced the subleading $v^6/r^{14}$ interaction . 

On the supergravity side, the D0-D0 scattering amounts to studying 
the D0-brane probe action in the presence of background fields corresponding 
to a D0-brane source and obtained by null reduction from eleven 
dimensions \cite{BBPT,Nfin}. 
This action gives predictions for terms of the 
form $v^{2{l+1}}/r^{7l};\ l=0,1,2,\ldots$ and reads
$$
   S_{0} = -T_0 \int \! dt \ h^{-1}[\sqrt{1-hv^2}-1] \ , 
$$
where $h=Q_0/r^7$, and $T_0,Q_0$ are the D0-brane's mass and charge, 
respectively. To explicitly verify that $S$ is reproduced in MT, one 
would have to carry out the highly demanding computation of three loop and 
higher diagrams. In this work we adopt a different approach: we will assume 
that S is in fact properly reproduced, use this assumption to propose a
general form for the MT effective action, and then test our proposal by 
using the action to compute the interactions of other types of branes. The 
point is that the same action which leads to $S_{0}$ when the background 
describes D0-branes must, by consistency, also yield the correct action 
for, say, a D2-brane and a D0-brane, when the background is changed 
accordingly. These consistency restrictions are so demanding that they 
plausibly prescribe a unique form for the effective action governing a 
wide class of backgrounds. 

The sorts of backgrounds we are interested in, those describing two 
separated objects, are given by the matrices
$$
   X_i = \left( \begin{array}{cc}
     (\Ucal_i )_{N_1 \times N_1} & \mbox{} \\
 \mbox{} & (\Vcal_i )_{N_2 \times N_2} \end{array} \right) \ \ ; \ \ 
i=1\ldots 9
$$
where $\Ucal_i,\Vcal_i$ are themselves matrices of variable size. To write 
our ansatz for the effective action we need a few definitions. We define a 
field strength tensor $\Fcal_{MN}$ by:
\begin{eqnarray}
 \Fcal_{0i} &=& \dot{\Ucal}_i \otimes \unitm 
 - \unitm \otimes \dot{\Vcal}^*_i  \label{e1} \\
 \Fcal_{ij} &=& -i[\Ucal_i,\Ucal_j]\otimes \unitm 
 + \unitm \otimes i[\Vcal^*_i,\Vcal^*_j] \label{e2}
\end{eqnarray}
Next, we define a ``harmonic function'' $\hsf$ by
\be
  \hsf = Q_0 \ \left(\, [\Ucal_i\otimes \unitm - \unitm \otimes \Vcal^*_i]
    [\Ucal_i\otimes \unitm - \unitm \otimes \Vcal^*_i]\, \right)^{-7/2} \ \ .
\label{e3}
\ee
With these definitions in hand, our proposal for the effective action reads
\be
    S_{MT} = -T_0\, \Tr \int \! dt\ \hsf^{-1} \{ \sqrt{-\det [\eta_{MN}
            - \hsf^{1/2} \Fcal_{MN}]} - 1 \} \ \ ,
\ee
where $\eta_{MN} = {\rm diag} (1,-1,-1,\ldots ,-1)$, and $\Tr$ means a 
particular trace operation over $\Ucal ,\Vcal$ space to be discussed in the 
next section. It is simple to check (see section 3) that $S_{MT}$ 
reduces to $S_{0}$ when $\Ucal_i,\Vcal_i$ are chosen to be 
D0-brane backgrounds. We should emphasize that $S_{MT}$ is not meant to be 
the full effective 
action of Matrix Theory, but rather a portion of it which is valid under 
restricted conditions. For instance, important simplifications occur when 
the field strength commutes:
$$
  [\Fcal_{\mu \nu} , \Fcal_{\alpha \beta} ] 
  = [\Fcal_{\mu \nu} , \Ucal_i \otimes \unitm ] 
  = [\Fcal_{\mu \nu} , \unitm \otimes \Vcal_i ] = 0 \  \ ;
$$
we will refer to this case as ``abelian''. If the background is 
not abelian in this sense, there will be corrections to $S_{MT}$ 
depending on commutators. (See \cite{NABI} for a related discussion). 
Nevertheless, $S_{MT}$ is sufficiently general 
that it encompasses a wide range of source-probe type calculations. 

The remainder of this paper is devoted to motivating and testing $S_{MT}$ for 
a variety of backgrounds. In section 2 we review some basic principles and 
establish conventions. Our ansatz for $S_{MT}$ is motivated in section 3 by 
recalling the cases where the loop diagrams have been computed 
explicitly, and demanding consistency with those results. We perform a number 
of detailed checks of $S_{MT}$ in section 4 by studying the action of 
a D2-brane probe in the presence of a D0-brane source. We consider
 rotating the D2-brane and show that the action changes in the correct 
way to agree with supergravity. In section 5 we consider replacing the 
D2-brane probe by a general D2p-brane probe, and show that the agreement 
with supergravity persists. We also include transverse velocities and
worldvolume electric fields on 
the probe. In section 6  we replace 
the D0-brane source by a D4-brane source. For reasons to be discussed, our 
checks in this case are restricted to the one loop level. We study the case 
of a D0-brane probe as well as a D4-brane probe oriented at an angle 
relative to the source. Again, we find agreement. In section 7 we review 
our results and discuss the prospects for extending our methods to more 
general backgrounds. Finally, an Appendix contains the derivations of 
D0-brane and D4-brane supergravity backgrounds via null reduction from 
eleven dimensions.

\section{Conventions}

In string units (we set $2\pi \alpha' = 1$) the Matrix Theory action 
is \cite{BFSS}
\be
 S = \frac{T_0}{2} \int \ dt \ \Tr \{ (D_tX_i)^2 + \half [X_i,X_j]^2   
   + 2 \Theta^T \dot{\Theta} - 2\Theta^T \gamma_i [\Theta ,X^i] \}
\ee
where $i,j=1,\ldots ,9$, $D_t = \pat_t - i[A_t, \cdot ]$ and $T_0$ is the
mass of a D0-brane. We will be considering various  backgrounds 
for the bosonic fields $X_i$. According to Susskind's DLCQ 
proposal \cite{Nfin}, 
when the $X_i$ are $\nbyn$ matrices, one is studying M theory with a null 
direction compactified and with $N$ units of longitudinal momentum. 

The background corresponding to $N$ D0-branes is 
$$
   X_i = \left( \begin{array}{cccc} x^{(1)}_i & \mbox{} & \mbox{} & \mbox{} \\
         \mbox{} & x^{(2)}_i & \mbox{} &  \mbox{} \\
         \mbox{} & \mbox{} & \ddots & \mbox{}  \\
         \mbox{} & \mbox{} & \mbox{} & x^{(N)}_i \end{array} \right) 
$$
where $x^{(a)}_i$ represents the position of the $a$th D0-brane. 

To construct the background corresponding 
to a D2-brane \cite{DeWetal,BFSS}, one introduces the 
canonical variables $Q,P$ satisfying $[Q,P]=i$, which are thought of as 
matrices of infinite size. The trace operation over $Q,P$ space is given by
\be
\Tr ~ \rightarrow ~ \frac{1}{2\pi}\int \! dP\,dQ.
\label{e6}
\ee
  In Matrix Theory, one 
considers not a pure D2-brane, but rather a bound state of D0-branes and 
a D2-brane -- the so called (2+0) configuration. The density $\sigma_0$ of 
D0-branes on the D2-brane is described in terms of a magnetic 
field \cite{Town,D}: $\sigma_0 = F_{12}/2\pi$, where we take the 
brane to lie in 
the (12)-plane. The MT background for such a configuration is 
\be
  X_1 = \frac{Q}{\sqrt{F_{12}}} \ ; \ X_2 = \frac{P}{\sqrt{F_{12}}}
        \ ; \ X_{I>2} = 0 \ . 
\ee
The division by $\sqrt{F_{12}}$ gives the D2-brane the correct central 
charge in the D=11 SUSY algebra \cite{BSS}. The above background represents 
the ``minimal'' (2+0) configuration, to which one can add electric fields, 
boosts, rotations, and non-extremal excitations. In subsequent sections we'll 
describe how these variables are represented in Matrix Theory. 

Other Dp-branes are represented by essentially repeating the above 
procedure. For instance, a four brane in the (1234)-hyperplane is given by
\be
 X_1 = \frac{Q_1}{\sqrt{F_{12}}} \ ; \ X_2 = \frac{P_1}{\sqrt{F_{12}}} 
\ ; \  X_3 = \frac{Q_2}{\sqrt{F_{34}}}
\ ; \ X_4 = \frac{P_2}{\sqrt{F_{34}}} \ .
\ee
This object is a (4+2+2+0) configuration consisting of a 
D4-brane, D2-branes in the (12)-plane with density $F_{12}/2\pi$, D2-branes 
in the (34)-plane with density $F_{34}/2\pi$, and D0-branes with 
density $F_{12}F_{34}/(2\pi )^2$.
One can also construct D4-brane configurations with vanishing D2-brane 
density \cite{HT}, as we discuss in section 6. 

Now let us turn to the supergravity description of D-brane interactions. A 
source D0-brane is described by the field configuration
\begin{eqnarray}
 ds^2 &=& h^{1/2} dt^2 - h^{-1/2} (dx^2_1 + \cdots + dx^2_9 ) \nonumber \\
 e^{-\phi} &=& h^{-3/4} \ \ ; \ \ C^{(1)}_t = h^{-1} \label{e9} \\
 h &=& \frac{Q_0}{r^7} \ . \nonumber
\end{eqnarray}
The solution above is not the standard D0 solution of IIA supergravity, but 
instead comes from the null reduction of a graviton solution in D=11, as 
reviewed in the Appendix. The primary difference with the standard solution 
is that normally $h=1+(Q_0/r^7)$ but here the 1 is absent. As discussed 
in \cite{BBPT}, this form for $h$ is crucial to obtain agreement with 
Matrix Theory. 

The interaction of the D0-brane with a Dp-brane will be described by 
treating the Dp-brane as a probe. The action for the probe is the 
Born-Infeld action with a Chern-Simons term \cite{Leigh,D}:
\be
 S_p = -T_p \int \! d^{p+1}\xi \ \left\{  e^{-\phi} 
 \sqrt{\det [g_{MN}\pat_{\mu} X^M 
    \pat_{\nu} X^N - F_{\mu \nu}]} 
   - \ \frac{1}{2^{p/2}(p/2)!} \epsilon_{i_1\cdots i_p} 
     F_{i_1i_2} \cdots F_{i_{p-1}i_p} C^{(1)}_t \right\} \ ,
\label{e10}
\ee
where $T_p = T_0/(2\pi)^{(p/2)}$ is the tension of the $p$-brane. The indices
$M,N$ are spacetime indices in 10 dimensions, the indices $\mu ,\nu$ are 
indices in the $p+1$ dimensional worldvolume, and the 
indices $i_1,i_2,\ldots$ are
spacelike indices in the worldvolume. The above probe action is valid for 
fields which are slowly varying (on a length scale set by the string scale) 
on the worldvolume; for rapidly varying fields one expects derivative 
corrections. By plugging in the background fields of the D0-brane one 
obtains an effective action for the Dp-brane which governs its interaction 
with the D0-brane to all orders in the string coupling. One of our 
goals is to show how such an action can arise from Matrix Theory.

\section{The general action for two-body interactions}

Let us consider the Matrix Theory (MT) background for two arbitrary
separated objects
\be
 (X_i)_{(N_1 + N_2)\times (N_1 + N_2)} = \left( \begin{array}{cc}
     (\Ucal_i )_{N_1 \times N_1} & \mbox{} \\
      \mbox{} & (\Vcal_i )_{N_2 \times N_2} \end{array} \right) \ .
\ee
Interactions between objects $\Ucal$ and $\Vcal$ arise by expanding the 
Matrix Theory action around this background and integrating out the massive 
degrees of freedom to a given number of loops. In general, one expects to 
obtain a horribly complicated non-linear function 
of $\Ucal_i$ and $\Vcal_i$. However, if attention is restricted to a 
subset of the diagrams, we'll argue that the terms sum up to give a simple 
form for the action. To motivate this formula let us first review the cases 
in which the loop calculations have been performed explicitly. 

First consider the interaction of two D0-branes \cite{Bachas,DFS,KP}. 
We take $\Ucal_i = (b_i + v_i t)\cdot \unitm_{N_1 \times N_1}$;
$\Vcal_i = \zerom_{N_2 \times N_2} $. In \cite{BBPT} it was shown that the 
two loop effective action coincides with the result of expanding
\be
 S_{0} = N_1 T_0 \int \!dt \,h^{-1} \ [\sqrt{1-hv^2} -1] \ \ ;  \ \ 
h=\frac{N_2Q_0}{b^7}
\label{e12}
\ee
to order $h^2$. $S_{0}$ is the action of a probe D0-brane moving in the 
background of a D0-brane source. The loop expansion of Matrix Theory 
corresponds to the $h$
expansion of supergravity, so if MT is correct, the infinite series of 
higher loop diagrams should reproduce the full form for $S_{0}$. It is of 
course important to check explicitly that this is indeed the case, but here 
we will simply assume that it is true. 

In fact the MT action contains many more terms than those displayed 
in (\ref{e12}) \cite{BBPT,BM}. 
An important point concerns the $N_1$ and $N_2$ dependence of
$S_{0}$; specifically, $S_{0}$ is linear in $N_1$. In general, an $l$
 loop diagram 
yields terms of the form $N^{\alpha}_1 N^{\beta}_2$ 
with $\alpha + \beta = l+1$. However, the source-probe calculations in 
supergravity that we are comparing with only correspond to terms linear 
in the number of D0-branes of the probe, thus we keep only those terms in MT 
as well. In a similar vein, the $v$ and $b$ dependences of the full MT 
action will differ from what appears in $S_{0}$. In particular, the $l$ loop 
diagrams can yield $b^{4-3l}f(v^2/b^4)$ whereas $S_{0}$ yields only terms 
like $v^{2(l+1)}/b^{7l}$. The extra terms in MT correspond to effects 
not included in our supergravity calculations -- such as the presence 
of ${\cal R}^4$ terms \cite{BM,RT,GGV} in 
the gravitational action -- so we drop them. 
To summarize: by keeping terms in the MT action which are linear in $N_1$ 
and of the form $v^{2(l+1)}/b^{7l}$, we expect to find complete agreement 
with the probe action from supergravity. 

To obtain more information we turn to backgrounds which are more general, but 
for which only one loop results are known. Ref. \cite{CT1} considered the 
background
\begin{eqnarray}
X_i &=& \left( \begin{array}{cc}
     (\Ucal_i )_{N_1 \times N_1} & \mbox{} \\
 \mbox{} & (\Vcal_i )_{N_2 \times N_2} \end{array} \right) \ \ i=1\ldots 7 \\
X_8 &=& \left( \begin{array}{cc}
     b\cdot \unitm_{N_1 \times N_1} & \mbox{} \\
 \mbox{} & \zerom_{N_2 \times N_2} \end{array} \right) \\
X_9 &=& \left( \begin{array}{cc}
     vt\cdot \unitm_{N_1 \times N_1} & \mbox{} \\
 \mbox{} & \zerom_{N_2 \times N_2} \end{array} \right)
\end{eqnarray}
and evaluated the one-loop determinants for arbitrary static and ``abelian'' 
$\Ucal_i,\Vcal_i$.  The determinants can be converted into an effective action,
which reads:
\be
S_{1 \,  \rm{loop}}= -\frac{1}{8}T_{0}\, \Tr^{(N_1)}\, \Tr^{(N_2)}  
\int \! dt \ \hsf \left[  \tr (\eta \Fcal )^4 
  - \frac{1}{4}(\tr (\eta \Fcal )^2 )^2 \right]
\label{e16}
\ee
Here, $tr$ refers to a trace over Lorentz indices.

We wish to write down an ansatz for the MT effective action which 
reproduces the special cases just described. Given this constraint, 
there is a very natural guess for the action:
\be
  S_{MT} = -T_0\, \Tr^{(N_1)}\, \hat{\Tr}^{(N_2)} 
     \int \! dt \ \hsf^{-1} \{ \sqrt{-\det [\eta_{MN} 
        - \hsf^{1/2} \Fcal_{MN} ]} - 1 \} \ \ .
\label{e17}
\ee
The $\hat{\Tr}^{(N_2)}$ operator requires some explanation, which we provide 
with reference to double line notation.  In this notation, an $l$ loop planar 
diagram contributing to the action consists of an outer loop associated with 
an index in $\Ucal_i$ space, and $l$ closed inner loops associated with an 
index in $\Vcal_i$ space. Each loop gives rise to a trace in the 
corresponding space, so an $l$ loop diagram 
gives, schematically, $\Tr^{(N_1)} (\Tr^{(N_2)})^l$. Each loop contains 
a number of insertions of the background field and these appear inside 
the traces. However, there are various ways of partitioning the background 
field operators among the various loops and this leads to a number of 
distinct diagrams at a given loop order. Consider, for example, the two 
loop (order $\hsf^2$) contribution:
\be
 S_{{\rm 2-loop}} \sim \Tr^{(N_1)}\Tr^{(N_2)}\Tr^{(N_2)} \hsf^2 \Fcal^6 \ \ .
\ee
This expression could stand for (schematically)
\be
 \Tr^{(N_1)} \{ \Tr^{(N_2)}(\hsf^2\Fcal^6)\Tr^{(N_2)}(\unitm_{N_2 \times N_2}) 
              + \Tr^{(N_2)}(\hsf \Fcal^4)\Tr^{(N_2)}(\hsf \Fcal^2) + \cdots 
             \} \ \ .
\ee
Without explicitly evaluating the diagrams, there is an ambiguity as to the 
relative weighting of the different terms, and the ambiguity only gets worse 
for higher orders in $\hsf$.  Fortunately, there are two important cases 
for which the ambiguity is absent. The first case arises when we consider 
D0-brane sources: $\Vcal_i = \zerom_{N_2 \times N_2}$. In such cases $\hsf$ 
and $\Fcal $ are proportional to $\unitm_{N_2 \times N_2}$ and so it does 
not matter how the traces are distributed: every way gives $N^l_2$. Thus 
for D0-brane sources the $\hat{\Tr}^{(N_2)}$ operator is equal to $N^l_2$ 
when acting on a term of order $\hsf^l$. Therefore, for D0-brane sources, we 
should be able to compare to all orders our ansatz for the action against 
the supergravity prediction for an arbitrary brane -- extremal or 
non-extremal -- moving in the background of a D0-brane. We will perform 
these checks in the next sections. 

The other unambiguous case occurs when we restrict attention to 1-loop 
results. 
In this case $\hat{\Tr}^{(N_2)} \rightarrow \Tr^{(N_2)}$, {\em i.e.} just 
a single standard trace. This time we can consider other sources besides 
D0-branes, and the results should still match with supergravity. We check 
this in section 6 by considering D4-brane sources. 
At the one loop level our proposed action can be proven directly by evaluating
the relevant Feynman diagrams, but at this time this has only been done for
various special cases.  
Note that aspects of the one loop correspondence between supergravity and MT 
have previously been discussed in \cite{CT2}.

As a first test of our 
ansatz, we will verify that it indeed reproduces (\ref{e12}) and (\ref{e16}). 
First consider 
the D0-D0 case. In this example
\be
\Fcal_{0i} = v_i \ ; \ \Fcal_{ij} = 0 \ ; \
      \hsf = \frac{Q_0}{|\vec{b} + \vec{v}t|^{7}}
\ee
and
\be
\det [\eta_{MN} - \hsf^{1/2} \Fcal_{MN}] = -(1-\hsf v^2) \ .
\ee
So,
\begin{eqnarray*}
 S_{MT} &=& - T_0 \Tr^{(N_1)} \hat{\Tr}^{(N_2)} \int \! dt \ \hsf^{-1}
 \{ \sqrt{1-\hsf v^2} - 1 \} \\
       &=& -N_1 T_0 \int \! dt \ (N_2 \hsf)^{-1} 
 \{ \sqrt{1 - N_2\hsf v^2} - 1 \} \\
       &=& -N_1 T_0 \int \! dt \ h^{-1} \{ \sqrt{1 - h v^2} - 1 \} \ \ , 
\end{eqnarray*}
as desired. The other special case (\ref{e16}) follows 
immediately upon expanding to 
order $\hsf$:
$$
 \hsf^{-1} \sqrt{-\det [ \eta_{MN} - \hsf^{1/2} \Fcal_{MN}]}
 = {\cal O}(\hsf^0) + \frac{\hsf}{8} \left[ \tr (\eta \Fcal )^4 
  - \frac{1}{4}(\tr (\eta \Fcal )^2 )^2 \right] + {\cal O} (\hsf^2) \ .
$$
Now we turn to more demanding checks of the ansatz.

\section{$0_s$ -- $(2+0)_p$ brane interactions}

In this section we will check our ansatz for the Matrix Theory action within 
the context of the D0-brane, D(2+0)-brane system. After treating the simplest
case, we will study the effect of rotating the (2+0) brane and recover
actions which agree with supergravity to all orders. 

The supergravity result which we will try to reproduce from Matrix Theory 
comes from the Born-Infeld action for the (2+0) probe.  We use the action
(\ref{e10}) with $p=2$:
\be
 S_2 = -\frac{T_0}{2\pi} \int \! d^3\xi \ \left\{ e^{-\phi}
        \sqrt{\det [g_{MN} \pat_{\mu} X^M \pat_{\nu} X^N -F_{\mu \nu}] }
        - F_{12}C^{(1)}_t \right\} \ \ ,
\label{e22}
\ee
where $g_{MN},e^{-\phi},C^{(1)}_t$ are the fields of a D0-brane 
given in (\ref{e9}). 

\subsection{Simplest case}

We first consider the simplest case of a stationary brane with vanishing
electric fields $F_{0i}$.  We will choose the static gauge for the probe, which
consists of setting the worldvolume coordinates $\xi^{0,1,2}$ to be equal to
the spacetime coordinates $X^{0,1,2}$,
\be
X^{0,1,2} = \xi^{0,1,2}  \ \ ; \ \ X^3 = b_3 
\ \ ; \ \ X^{i>3} = 0 .
\ee
The worldvolume gauge field is taken to have only magnetic components:
\be
\half F_{\mu \nu} \ d\xi^\mu \wedge d\xi^\nu 
         = F_{12}\  d\xi^1 \wedge d\xi^2. 
\ee
Inserting these fields into (\ref{e22}) yields for the probe action,
\be
S_2 = -\frac{T_0}{2\pi} \int \! d^3\xi \ h^{-1} 
       \{ \sqrt{F^2_{12}+h} - F_{12} \}.   \label{e25} 
\ee
The harmonic function $h$ is given in (\ref{e9}), where $r$ represents 
the spacetime
separation between the D0-brane and a point on the (2+0) worldvolume.  So,
\be
h = N_2Q_0\ [(\xi^1 )^2 + (\xi^2 )^2 + (b_3 )^2 ]^{-7/2}  .
\label{e26}
\ee

Now we turn to Matrix Theory to see if our ansatz can reproduce (\ref{e25}). 
In MT we represent we represent a (2+0) state lying in the (12) plane by
\begin{eqnarray*}
\Ucal_1 &=& \frac{Q}{\sqrt{F_{12}}}  \ \ \ ;  \ \ \ 
\Ucal_2  = \frac{P}{\sqrt{F_{12}}} \\
\Ucal_3 &=& b_3  \ \ \ ;  \ \ \ U_{i>3} = 0  \ \ \ ;  \ \ \ 
\Vcal_i = \zerom_{N_2 \times N_2} \ .
\end{eqnarray*}
The MT field strength, as defined in (\ref{e1}),(\ref{e2}), is then
\be
\half \Fcal_{MN} dX^{M} \wedge dX^{N} 
    = -i\frac{[Q,P]}{F_{12}} dX^1 \wedge dX^2 
    = \frac{1}{F_{12}} dX^1 \wedge dX^2 
\ee
The ``harmonic function'' $\hsf$, defined by (\ref{e3}), becomes
\be
\hsf = Q_0 \ \left(\frac{Q^2}{F_{12}} + 
\frac{P^2}{F_{12}} + b^2_3 \right)^{-7/2} 
\ee
Working out the determinant in $S_{\rm{MT}}$ we find
\be
\det[\eta_{MN} - \hsf^{1/2} \Fcal_{MN}] 
     = -\left(1+\frac{\hsf }{F^2_{12}}\right). 
\ee
It remains to compute the traces.  $P, Q$ become the worldvolume coordinates
when we convert the trace to an integral using (\ref{e6}) and rescale,
$$
 \frac{P}{\sqrt{F_{12}}} = \xi^1 \ ; \ \frac{Q}{\sqrt{F_{12}}} = -\xi^2
 \ \Rightarrow \ \Tr = \frac{1}{2\pi } \int \! dQ \,dP 
  = \frac{F_{12}}{2\pi } \int \! d^{2}\xi
$$
Note that the integration measure is implicitly a two-form, which explains 
the positive sign for the $d^{2}\xi$ measure. 
$\Tr^{(N_{2})} $ simply acts by multiply $\hsf$ by $N_{2}$.  Putting the pieces
together and substituting into (\ref{e17}) yields for the effective action
\be
S_{MT} = -\frac{T_0}{2\pi} \int \! d^3\xi \ h^{-1} 
       \{ \sqrt{F^2_{12}+h} - F_{12} \},
\ee
where
\be
h= N_2Q_0\ [(\xi^1 )^2 + (\xi^2 )^2 + (b_3 )^2 ]^{-7/2}
\ee
as in (\ref{e26}). The result indeed agrees with (\ref{e25}). 

\subsection{Rotated case}

As our next check, let's consider the effect of rotating the D2-brane in 
the (13)-plane. Since the D0-brane source background is spherically symmetric,
this rotation is rather trivial from a physical point of view; however, it
constitutes a non-trivial check of our ansatz and illustrates the 
general procedure for rotating branes in more elaborate examples. It will 
also illustrate a relation between different gauge choices in supergravity
actions, and the various ways of writing backgrounds in Matrix Theory. 

In supergravity, with the usual static gauge, the rotated (2+0)-brane
background is given by 
\be
 X^{0,1,2} = \xi^{0,1,2} \  \ ; \ \ X^3 = b_3 + \xi^1 \tan \theta \ .
\ee
Its action in the D0-background is 
\be
 S_2 = -\frac{T_0}{2\pi} \int \! d^3\xi \ h^{-1} \{ 
             \sqrt{h(1+\tan^2 \theta ) +F^2_{12}} -F_{12} \} \ ,
\label{e33}
\ee
where the harmonic function is
\be
h = N_2Q_0\ [(\xi^1 )^2 + (\xi^2 )^2 
     + (b_3 + \xi^1 \tan \theta )^2 ]^{-7/2}  \ \ .
\label{e34}
\ee
The action looks simpler in an alternative gauge, where the spatial 
worldvolume coordinates measure the physical distance along the brane.
This gauge choice is easily written with the help of rotated spacetime
coordinates $\Xbar^M$ which are aligned with the brane:
\begin{eqnarray}
\Xbar^M &=& X^M \ \  \ \ M \neq 1,3  \\
\left( \begin{array}{c} \Xbar^1 \\ \Xbar^3 -b_3 \end{array} \right)
&=&  \left( \begin{array}{cc} \cos \theta & \sin \theta \\ 
               -\sin \theta & \cos \theta \end{array} \right)
 \ \left( \begin{array}{c} X^1 \\ X^3 -b_3 \end{array} \right) \ .
\end{eqnarray}
We fix the new gauge, which we will call the ``aligned gauge'', by
setting 
\be
 \Xbar^{0,1,2} = \xibar^{0,1,2}.
\ee
In these coordinates (2+0)-brane is located at $\Xbar^3 = b^3$. 
The new worldvolume coordinates
are related to the old ones by 
\be
\xibar^0 = \xi^0  \ \ ;  \ \ \xibar^1 \cos \theta = \xi^1  \ \ ;  \\ \xibar^2 = \xi^2 \ .
\ee
Therefore, the new worldvolume magnetic field 
$F_{\barone \bartwo}$ is related to $F_{12}$ by
\be
     F_{\barone \bartwo} = \cos \theta F_{12} \ .
\ee
The D2-brane action is now
\be
 S_2 = -\frac{T_0}{2\pi} \int \! d^3\xibar \ \barh^{-1} 
       \{ \sqrt{F^2_{\barone \bartwo }+\barh} - F_{\barone \bartwo} \} \ \ ,
\label{e41}
\ee 
where the harmonic function is simply
\be
 \barh = N_2Q_0\ [(\xibar^1 )^2 + (\xibar^2 )^2 
     + (b_3 )^2 ]^{-7/2} \ .
\ee 
Naturally, the two forms of the action (\ref{e33}),(\ref{e41}) are 
related by the 
relations of the barred and unbarred variables. 

In the rotated frame, it is simple to write down the MT background:
\begin{eqnarray*}
\Ucal_{\barone} &=& \frac{Q}{\sqrt{F_{\barone \bartwo}}} 
 \ \ \ ; \ \ \  \Ucal_{\bartwo}  = \frac{P}{\sqrt{F_{\barone \bartwo}}} \\
\Ucal_{\bar{3}} &=& b_3 \ \ \  ; \ \ \ \Ucal_{\bar{i}>3} = 0 
                 \ \ \ ; \ \ \ \Vcal_{\bar{i}} = \zerom_{N_2 \times N_2} \ .
\end{eqnarray*}
The MT effective action reduces to the supergravity
action for the probe in the aligned gauge, and this will continue to 
hold true as
we return back to the static gauge. To make contact with the static gauge
supergravity action, we rewrite the MT background by rotating back to the
original unbarred coordinates. To do this correctly,
we note that $\Ucal_{\barone},\Ucal_{\bartwo}$ are related to the spacetime coordinates
$\Xbar^2,\Xbar^1$:
$$
 -\Ucal_{\barone} = \xibar^2 = \Xbar^2 \  \ ; \ \ \
 \Ucal_{\bartwo} = \xibar^1 = \Xbar^1 \ .
$$
Therefore, a spacetime rotation in the (13) plane  becomes a rotation 
in the (23)-plane of the $\Ucal$'s:
\be
\left( \begin{array}{c} -\Ucal_1 \\ \Ucal_2 \\ \Ucal_3 -b_3 \end{array} \right)
  = \left( \begin{array}{ccc} 1 & 0 & 0 \\
              0 & \cos \theta & -\sin \theta \\ 
              0 & \sin \theta & \cos \theta \end{array} \right)
 \ \left( \begin{array}{c} -\Ucal_{\barone} 
    \\ \Ucal_{\bartwo} \\ \Ucal_{\bar{3}} -b_3 \end{array} \right) \ .
\ee
So, the background given in terms of $\Ucal$'s is
\begin{eqnarray}
 \Ucal_1 &=& \Ucal_{\barone} = \sqrt{\frac{1}{\cos \theta F_{12}}} Q \\
 \Ucal_2 &=& \cos \theta \ \Ucal_{\bartwo} = \sqrt{\frac{\cos \theta}{F_{12}}} P \\
 \Ucal_3 &=& b_3 + \sin \theta \ \Ucal_{\bartwo} = b_3 
               + \tan \theta \sqrt{\frac{\cos \theta}{F_{12}}} P \ .
\end{eqnarray}
Thus, the MT field strength tensor is
\be
 \Fcal_{MN} = \left( \begin{array}{ccccc} 0 & 0 & 0 & 0 & \cdots \\
                    \mbox{} & 0 & F^{-1}_{12} & \tan \theta F^{-1}_{12} 
                           & \mbox{} \\
                    \ddots & \mbox{} & 0 & 0 & \mbox{} \\
                    \mbox{} & \ddots & \mbox{}  & 0 & \mbox{} \\
                    \mbox{} & \mbox{} & \mbox{} & \mbox{}  & \ddots
                  \end{array} \right) 
\label{e47}
\ee
and the harmonic function is
\be
 \hsf = Q_0 \ \left[\frac{Q^2}{\cos \theta F_{12}} 
    + \frac{\cos \theta \ P^2}{F_{12}} + (b_3 + \tan \theta 
        \sqrt{\frac{\cos \theta }{F_{12}}} P)^2 \right]^{-7/2} \ . 
\ee
Changing integration variables:
\be
\sqrt{\frac{\cos \theta}{F_{12}}} P = \xi^1 \ ; \
\sqrt{\frac{1}{\cos \theta F_{12}}} Q = -\xi^2 
\ \Rightarrow  \ \Tr = \frac{1}{2\pi } \int \! dQ\,dP 
  = \frac{F_{12}}{2\pi } \int \! d^{2}\xi .
\ee
The harmonic function now takes the same form (\ref{e34}) as in 
the supergravity
calculation. Also, inserting the field strength (\ref{e47}) into 
the MT effective
action (\ref{e17}), we recover the supergravity action (\ref{e33})
in  static gauge.

The preceeding discussion was quite detailed, but 
relations like the ones above  are needed when one moves on
to consider the more complicated examples of non-extremal, boosted, and
rotated brane configurations.

\subsection{Two rotations}

As our next example, we add another rotation. We take the D2-brane to 
be rotated by the angle $\theta_1$ in (13)-plane and by the angle 
$\theta_2$ in (24)-plane.  The two coordinate systems are  related by
\be
\left( \begin{array}{c} \Xbar^1 \\ \Xbar^2 \\ \Xbar^3 -b_3 
        \\ \Xbar^4 -b_4 \end{array} \right)
  = \left( \begin{array}{cccc}\cos \theta_1  & 0 & \sin \theta_1 & 0 \\
              0 & \cos \theta_2 & 0 & \sin \theta_2  \\ 
              -\sin \theta_1 & 0  & \cos \theta_1 & 0 \\
         0 & -\sin \theta_2 & 0 & \cos \theta_2 \end{array} \right)
 \ \left( \begin{array}{c} X^1
    \\ X^2 \\ X^3 -b_3 \\ X^4 -b_4 \end{array} \right) \ .
\ee
The two  ways to describe the brane are then
\begin{eqnarray}
X^{1,2} &=& \xi^{1,2} \ \ ; \ \ X^3 = \tan \theta_1 \xi^1 + b_3
         \ \ ; \ \ X^4 = \tan \theta_2 \xi^2 + b_4  \ \ \ \ 
        {\rm (static)} \\
\Xbar^{1,2} &=& \xibar^{1,2} \ \ ; \ \ \Xbar^3 = b_3
      \ \ ;  \ \ \Xbar^4 = b_4 \ \ \ \ {\rm (aligned)}  \ .
\end{eqnarray}
The supergravity probe action is
\begin{eqnarray}
 S &=&  -\frac{T_0}{2\pi} \int \! d^3\xibar \ \barh^{-1} 
       \{ \sqrt{\barh + F^2_{\barone \bartwo}} - F_{\barone \bartwo} \} 
     \ \ \ {\rm (aligned )} \\
   &=& -\frac{T_0}{2\pi} \int \! d^3\xi \ h^{-1} 
       \{ \sqrt{h(1+\tan^2 \theta_1 )(1+\tan^2 \theta_2 ) + F^2_{12}} 
        - F_{12} \} \ \ {\rm (static)} \label{e54} 
\end{eqnarray}
where the harmonic function is
\begin{eqnarray}
 \barh &=& N_2Q_0\ [(\xibar^1 )^2 + (\xibar^2 )^2 
     + (b_3 )^2 + (b_4 )^2]^{-7/2}  \ \ \ \ {\rm (aligned)} \\
    h  &=& N_2Q_0\ [(\xi^1 )^2 + (\xi^2 )^2 
     + (b_3 +\xi^1 \tan \theta_1 )^2 
   + (b_4 +\xi^2 \tan \theta_2 )^2]^{-7/2}  \ \ \ {\rm (static)} . \label{e56}
\end{eqnarray}
In the barred variables, the MT background is 
\be
 \Ucal_{\barone} = \frac{Q}{\sqrt{F_{\barone \bartwo}}} 
 \ \ ; \ \ \Ucal_{\bartwo} = \frac{P}{\sqrt{F_{\barone \bartwo}}} 
 \ \ ;  \ \ \Ucal_{\bar{3},\bar{4}} = b_{3,4} \ \ .
\ee
The worldvolume coordinates and magnetic fields are related by
\be
   \xi^0 = \xi^0  \ \ ;  \ \ \xi^{1,2} = \xibar^{1,2} \cos \theta_{1,2} 
   \ \ ;  \ \ F_{\barone \bartwo} = \cos \theta_1 \cos \theta_2 F_{12} \ .
\ee
Rotating to the original frame, the (13) and (24) rotations correspond
to the (23) and (14) rotations for the $\Ucal$'s :
\be
\left( \begin{array}{c} -\Ucal_1 \\ \Ucal_2 \\ \Ucal_3 -b_3 
        \\ \Ucal_4 -b_4 \end{array} \right)
  = \left( \begin{array}{cccc}\cos \theta_2  & 0 & 0 & -\sin \theta_2 \\
              0 & \cos \theta_1 & -\sin \theta_1 & 0  \\ 
              0 & \sin \theta_1 & \cos \theta_1 & 0 \\
         \sin \theta_2 & 0 & 0  & \cos \theta_2 \end{array} \right)
 \ \left( \begin{array}{c} -\Ucal_{\barone}
    \\ \Ucal_{\bartwo} \\ \Ucal_{\bar{3}} -b_3 \\ \Ucal_{\bar{4}} -b_4 
       \end{array} \right) \ .
\ee  
This yields 
\begin{eqnarray}
 \Ucal_1 &=& \cos \theta_2 \ \Ucal_{\barone} 
        = \sqrt{\frac{\cos \theta_2}{\cos \theta_1 F_{12}}} \ Q \\
 \Ucal_2 &=& \cos \theta_1 \ \Ucal_{\bartwo} 
        = \sqrt{\frac{\cos \theta_1}{\cos \theta_2 F_{12}}} \ P \\
 \Ucal_3 &=& b_3 + \sin \theta_1 \ \Ucal_{\bartwo} = b_3 
               + \tan \theta_1 \ \Ucal_2 \\
 \Ucal_4 &=& b_4 - \sin \theta_2 \ \Ucal_{\barone} = b_4 
               - \tan \theta_2 \ \Ucal_1 \ \ ,
\end{eqnarray}
so the MT field strength $\Fcal_{MN} $ is related 
to the probe field strength $F_{\mu \nu}$ by
\be
  \Fcal_{MN} = \left( \begin{array}{ccccccc} 0 & 0 & 0 & 0 & 0 & 0 & \cdots \\
 \mbox{} & 0 & F^{-1}_{12} & \tan \theta_1 F^{-1}_{12} & 0 & 0 & \mbox{} \\
 \ddots & \mbox{} & 0 & 0 & \tan \theta_2 F^{-1}_{12} & 0 & \mbox{} \\
 \mbox{} & \ddots & \mbox{}  & 0 & -\tan \theta_1 \tan \theta_2 F^{-1}_{12} 
  & 0 & \mbox{} \\
 \mbox{} & \mbox{} & \ddots & \mbox{} & 0 & 0 & \mbox{} \\
 \mbox{} & \mbox{} & \mbox{} & \mbox{} & \mbox{} & \ddots & \mbox{} \\
 \mbox{} & \mbox{} & \mbox{} & \mbox{} & \mbox{} & \mbox{} & \mbox{} \\
                  \end{array} \right)  \ .
\label{e64}
\ee
Notice the component $\Fcal_{34}$, which has a quadratic dependence on the
slopes; it appears since $\Ucal_3,\Ucal_4$ above do not commute.

The harmonic function $\hsf$ in the MT effective action is
\be
\hsf = N_2 Q_0 \ \{ \Ucal^2_1 + \Ucal^2_2 
      + (b_3 + \tan \theta_1 \ \Ucal_2 )^2 
      + (b_4 - \tan \theta_2 \ \Ucal_1 )^2 \}^{-7/2} \ .
\ee
With the change of variables
\be 
 \Ucal_2 = \sqrt{\frac{\cos \theta_1}{\cos \theta_2 F_{12}}} \ P = \xi^1
 \ ; \ \Ucal_1 = \sqrt{\frac{\cos \theta_2}{\cos \theta_1 F_{12}}} \ Q = -\xi^2
 \ \Rightarrow \ \Tr = \frac{1}{2\pi } \int \! dQ\,dP 
  = \frac{F_{12}}{2\pi } \int \! d^{2}\xi
\ee
$\hsf$ becomes
\be
\hsf = N_2 Q_0 \ \{ (\xi^1 )^2 + (\xi^2 )^2 
      + (b_3 + \xi^1 \tan \theta_1 )^2 
      + (b_4 + \xi^2 \tan \theta_2 )^2 \}^{-7/2} \ ,
\label{e67}
\ee
in agreement with the harmonic function $h$ in supergravity (\ref{e56}).
Substituting in $\hsf$ (\ref{e67}) 
and $\Fcal_{MN}$ (\ref{e64}), the MT effective action (\ref{e17})
reduces to the supergravity action (\ref{e54}) for the 
probe in the static gauge.

We now move on to more general probes, and add electric fields and transverse
velocities to the backgrounds.

\section{$0_s$ -- $(p+(p-2)+\cdots +0)_p$ brane interactions}

In this section, we consider $p+(p-2)+\cdots +0$ brane probes (with $p=2n$)
in the 0-brane background. The results we find will generalize those of the
previous section and will highlight the remarkable way in which Matrix Theory
can accomodate a wide variety of branes.  To check our ansatz in a more 
detailed manner than before, we will    
turn on the electric components $F_{0i}$ of the worldvolume gauge field
$F_{\mu \nu}$ in addition to the magnetic components. We will show 
that the MT effective action ansatz (\ref{e17}) reproduces 
the supergravity action for the general p-brane probe.  
 Finally, we return to the example of the  (2+0)-brane probe,
 adding both transverse velocities and rotations, and demonstrate 
the resulting agreement with supergravity. 

Once again, we start with the supergravity action for the probe,
\be
 S_p = -T_p \int \! d^{p+1}\xi \ \left\{ h^{-3/4} \sqrt{\det [g_{\mu \nu}
                                                    - F_{\mu \nu}]}
    \ - \ h^{-1} {\rm Pf} (-B) \right\} \ .
\label{e68}
\ee
where we use the notation $B_{ij}$ for the spatial part of the worldvolume
field strength of the $2n+2(n-1)+\cdots +0$ brane probe:
\be
 (B_{ij}) = (F_{ij}) = \left( \begin{array}{ccccc}
        0 & F_{12} & \mbox{} & \mbox{} & \mbox{} \\
       -F_{12} & 0 & \mbox{} & \mbox{} & \mbox{} \\
       \mbox{} & \mbox{} & 0 & F_{34} & \mbox{} \\
       \mbox{} & \mbox{} & -F_{34} & 0 & \mbox{} \\
       \mbox{} & \mbox{} & \mbox{} & \mbox{} & \ddots \end{array} \right) \ ,
\ee
and we have made use of the Pfaffian:
\be
 {\rm Pf} (B) = \frac{(-1)^n}{2^{n}n!} \epsilon_{i_1\cdots i_{2n}} 
     B_{i_1i_2} \cdots B_{i_{2n-1}i_{2n}} \ .
\ee
In the square root term of the action, there appears determinant of the matrix
\be
 g-F = \left( \begin{array}{cc} h^{-1/2}(1-hv^2) &  -\bfE^T \\
             \bfE & -h^{1/2} \unitm -B \end{array} \right) \ ,
\ee
where we use the notation
\be
   \bfE = (E_i) = (F_{0i})
\ee
for the electric field components. We have also included a transverse
velocity $v$ in the direction $X^{p+1}$.
Next, we manipulate the square root of the
determinant by inserting a factor
\be
 1 = \sqrt{\det \left( \begin{array}{cc} 1 & \mbox{} \\
                                       \mbox{} & -B \end{array} \right)
   \det \left( \begin{array}{cc} 1 & \mbox{} \\
                                       \mbox{} & -B^{-1} \end{array} \right) }
   = {\rm Pf}(-B) \sqrt{ \det \left( \begin{array}{cc} 1 & \mbox{} \\
                                       \mbox{} & -B^{-1} \end{array} \right) } 
\ee
and pulling out a factor $h^{-1/4}$. Then the action (\ref{e68}) takes the form
\be
 S_p = -T_p \int \! d^{p+1} \xi \ {\rm Pf}(-B) \ h^{-1}
 \ \{ \sqrt{\det \bfM_1 } -1 \} \ ,
\label{e74}
\ee
where 
\be
  \det \bfM_1 \equiv \det 
     \left( \begin{array}{cc} 1-hv^2 &  -h^{1/2}\bfE^T \\
             -B^{-1}\bfE & \unitm + h^{1/2} B^{-1} \end{array} \right) \ .
\ee

We will now compare the above form of the supergravity action with the MT
effective action ansatz
\be
S_{MT} = -T_0  \Tr^{(N_1)}\int \! dt \ \hsf^{-1} 
 \left\{ \sqrt{ -\det [\eta - \hsf^{1/2} \Fcal ]} -1 \right\} \ .
\label{e76}
\ee
The $\Tr^{(N_2)}$ operation has already been performed, and its effect has
been
taken into account by including a factor $N_2$ into the ``harmonic function''
$\hsf$. Next, we map the action (\ref{e76}) to an 
action on the $p+1$-dimensional
worldvolume, using the rule
\be
  T_0 \Tr^{(N)} \int \! dt \rightarrow T_p \int \! d^{p+1}\xi \ F_{12}
                                                               F_{34} \cdots
     = T_p \int \! d^{p+1} \xi \ {\rm Pf}(-B) \ ,
\ee
and relating the matrix field strength $\Fcal$ to 
the $p$-brane worldvolume field strength $F$ and the 
transverse velocity $v$ by the identities
\begin{eqnarray}
\Fcal &=& \hat{g} \hat{F} \hat{g} \\
\hat{g} &=& {\rm diag} (1,\underbrace{-F^{-1}_{12},-F^{-1}_{12},
           -F^{-1}_{34},-F^{-1}_{34},\ldots }_{p},
        \underbrace{-1,\ldots,-1}_{9-p}) \\
\hat{F} &=& \left( \begin{array}{ccccc} 
       0 & \overbrace{F_{0j}}^p & v & 0 & \cdots \\
      -F_{0i} & F_{ij} & 0 & \mbox{} & \mbox{} \\
       -v & 0 & 0 & \mbox{} & \mbox{} \\
      0 & \mbox{} & \mbox{} & \ddots & \mbox{} \\
      \vdots & \mbox{} & \mbox{} & \mbox{} & \mbox{} \end{array} \right)  \ .
\label{e80}
\end{eqnarray}
These relations are the same as those used in previous sections, but written in
a different notation. 
The harmonic function $\hsf$ is reduced to the supergravity form $h$ as before.
With these substitutions, the MT effective action takes the form
\be
 S_{MT} = -T_p \int \! d^{p+1} \xi \ {\rm Pf}(-B) h^{-1} \ 
           \left\{ \sqrt{-\det [\eta - h^{1/2} \hat{g} \hat{F} \hat{g}]}
        - 1 \right\} \ .
\label{e81}
\ee
Obviously, this is equal to the supergravity action (\ref{e74}), iff 
the determinant
terms under the square roots  agree. To show this, we first rewrite
the determinant term in (\ref{e81}) as follows:
\be
 -\det [\eta - h^{1/2}\hat{g} \hat{F} \hat{g} ] 
 = \det [\unitm - h^{1/2} \eta \hat{g}^2 \hat{F} ] \ .
\ee
Substituting\footnote{Note that $\eta \hat{g}^2$ is equal to the metric
$\hat{\eta}$, the rescaling of the spatial coordinates, which was discussed
in Refs. \cite{PP,KVK}.} 
\be
 \eta \hat{g}^2 = {\rm diag } (1,-F^{-2}_{12},-F^{-2}_{12},\ldots ,
 -1,-1,\ldots ,-1) = \left( \begin{array}{ccc}
                 1 & \mbox{} & \mbox{} \\
                \mbox{} & B^{-2} & \mbox{} \\
         \mbox{} & \mbox{} & -\unitm \end{array} \right) 
\ee
and using the expression (\ref{e80}) for $\hat{F}$, we get
\be
 -\det [\eta - h^{1/2}\hat{g} \hat{F} \hat{g} ] = \det
     \left( \begin{array}{cccc} 1 & -h^{1/2}\bfE^T & -h^{1/2}v & 0 \\
           h^{1/2}B^{-2} \bfE & \ \unitm - h^{1/2} B^{-1} & 0 & \mbox{} \\
         -h^{1/2}v & 0 & 1 & \mbox{} \\
        0 & \mbox{} & \mbox{} & \unitm \end{array} \right) \ .
\ee
The $10\times 10$ determinant above is equal to the $(p+1)\times (p+1)$
determinant
\be
 \det \left( \begin{array}{cc} 1-hv^2 & -h^{1/2} \bfE^T \\
               h^{1/2} B^{-2} \bfE & \ \unitm - h^{1/2} B^{-1} 
        \end{array} \right) \equiv \det \bfM_2 \ .
\ee
Then, finally, the agreement of the MT effective action and the supergravity
effective action is implied by the following determinant identity
\be
     \det \bfM_2 = \det \bfM_1 \ ,
\ee
which we have verified by an explicit evaluation of the two determinants.

To conclude this section, we consider an example which combines the
results given above with the results of the previous section: we consider a
  2+0 brane with electric fields, rotated
in the (13) and (23) planes, and with a  velocity in direction 3.
In the static gauge, the background is given by
\be
 X^{0,1,2} = \xi^{0,1,2} \ ; \ X^3 = b_3 + vt + \xi^1 \tan \theta_1
                                         + \xi^2 \tan \theta_2 \ .
\ee
The (2+0$)_p$ action is 
\be
 S_2 = -T_2 \int \! d^{3} \xi \left\{ h^{-3/4} \sqrt{\det [g_{\mu \nu} 
     -F_{\mu \nu}]} - h^{-1}F_{12} \right\} \ ,
\label{e88}
\ee
where
\be
   (g_{\mu \nu} -F_{\mu \nu} ) = \left( \begin{array}{ccc}
      h^{1/2} (1-hv^2) & -h^{1/2} v \tan \theta_1 -F_{01}
           & -h^{1/2} v \tan \theta_2 - F_{02} \\
\ddots & -h^{1/2}(1 + \tan^2 \theta_1) 
     & -h^{1/2}\tan \theta_1 \tan \theta_2 -F_{12} \\
   \mbox{} & \mbox{} & -h^{1/2}(1+\tan^2 \theta_2 )
     \end{array} \right) 
\ee
and the harmonic function is 
\be
 h = N_1 Q_0 \ \{ (\xi^1)^2 + (\xi^2)^2 + (b_3 + \xi^1 \tan \theta_1
                 + \xi^2 \tan \theta_2 )^2 \}^{-7/2} \ .
\label{e90}
\ee
To find the MT background $\Ucal_i$ and the MT field strength, we proceed
as before: we first go to the aligned gauge, thus eliminating the rotation
contribution. Then we write down the MT background in the rotated frame
(recalling that the origin of the coordinates is set to the brane, as in
the previous section), and rotate back to the original frame. The result
is as follows:
\begin{eqnarray}
 \Ucal_1 &=& \sqrt{\Fcal_{12}} \ Q + \Fcal_{01} t  \nonumber \\
 \Ucal_2 &=& \sqrt{\Fcal_{12}} \ P + \Fcal_{02} t \\
 \Ucal_3 &=& \frac{\Fcal_{13}}{\sqrt{\Fcal_{12}}} \ P 
            -\frac{\Fcal_{23}}{\sqrt{\Fcal_{12}}} \ Q + b_3 +\Fcal_{03}t  
           \ , \nonumber
\end{eqnarray}
with the MT field strength $\Fcal_{MN}$ given by
\be
 \Fcal_{MN} = \left( \begin{array}{cccccc}
        0 & -F_{01}F^{-1}_{12} & -F_{02}F^{-1}_{12} 
          & -(\tan \theta_1 F_{02} - \tan \theta_2 F_{01})F^{-1}_{12} +v
          & 0 & \cdots \\
      \ddots & 0 & F^{-1}_{12} & \tan \theta_1 F^{-1}_{12} & 0 & \mbox{} \\
      \mbox{} & \ddots & 0 & \tan \theta_2 F^{-1}_{12} & 0 & \mbox{} \\
      \mbox{} & \mbox{} & \ddots & 0 & 0 & \mbox{} \\
      \mbox{} & \mbox{} & \mbox{} & \ddots & \mbox{} & \ddots 
     \end{array} \right) \ .
\ee
The MT harmonic function is
\be
 \hsf = N_2 Q_0 \ \{ \Ucal^2_1 
         + \Ucal^2_2 
         + \Ucal^2_3 \}^{-7/2} \ .
\ee
With the change of variables
\be
\Ucal_2 = \xi^1 \ ; \ \Ucal_1 = -\xi^2 
 \ \Rightarrow \ \Tr = \frac{1}{2\pi} \int \! dQ\,dP = 
                           \frac{F_{12}}{2\pi} \int \! d^{2}\xi 
\ee
the harmonic function becomes
\be
\hsf = N_2Q_0 \{ (\xi^1)^2 + (\xi^2)^2 + (b_3 + \xi^1 \tan \theta_1
           + \xi^2 \tan \theta_2 )^2 \}^{-7/2} \ , 
\ee
in agreement with the form (\ref{e90}). 
Further, substituting $\Fcal_{MN}$ from above
into the MT effective action (\ref{e17}), we correctly 
reproduce the supergravity
probe action (\ref{e88}). Thus, even this 
complicated check for non-extremal, boosted
and rotated branes passes the test.  

\section{D4-branes as sources}

We have seen that our ansatz correctly reproduces a wide variety of actions
for probes moving in the background of a D0-brane source.  We now replace the
D0-brane source by a bound state of D4-branes and D0-branes: the (4+0)
configuration.  This replacement adds a great deal of additional structure
to the probe actions and serves as a highly non-trivial check of our methods.
However, for reasons discussed in section 3 we will confine our attention
to one loop results; it is an important challenge to resolve the trace
ambiguities that would allow for checks to be made for higher loops.

In supergravity the (4+0) configuration is described by the fields \cite{ATM}
\begin{eqnarray*}
 ds^2 &=& h^{-1/2}H_{4}^{-1/2} dt^2 - h^{1/2}H_{4}^{-1/2} (dx^2_1 + \cdots + dx^2_4 )- h^{1/2}H_{4}^{1/2} (dx^2_5 + \cdots + dx^2_9 ) \\
 e^{-\phi} &=& h^{-3/4}H_{4}^{1/4} \ \ ; \ \ C^{(1)}_t = h^{-1} \ \  ; \ \
C^{(5)}_{t1234} = H_{4}^{-1}-1 \\
 h &=& \frac{\tilde{F}^2 N_{4}Q_{0}^{(4)}}{(2\pi)^2r^{3}_{\perp}}= 
\frac{\tilde{F}^2 N_{4}Q_{0}}{15r^{3}_{\perp}}
 \ \ \ ; \ \ \   
  H_{4}=1+N_{4}h_{4}=1+\frac{N_{4}Q_{4}}{r^{3}_{\perp}} = 
1+\frac{N_{4}Q_{0}}{15 r^{3}_{\perp}}  \ .
\end{eqnarray*}

We are following the notation in \cite{CT1}: $Q^{(4)}_0$ is the charge
density of a D0-brane ``smeared'' in four directions, $Q_4$ is the D4-brane
charge, and $N_4$ is the number of D4-branes.
The Matrix theory background for the (4+0) state is obtained by combining 
two (4+2+2+0) states in such a way as to cancel off the D2-brane charge.
Explicitly \cite{HT}:
$$
 \Vcal_{1} = \frac{1}{\sqrt{\tilde{F}}}  \left( \begin{array}{cc}
     \tilde{Q}_{1} & 0 \\
      0 & \tilde{Q}_{1} \end{array} \right) 
\ \ \ ; \ \ \
 \Vcal_{2} = \frac{1}{\sqrt{\tilde{F}}}  \left( \begin{array}{cc}
     \tilde{P}_{1} & 0 \\
      0 & -\tilde{P}_{1} \end{array} \right) 
$$
$$
 \Vcal_{3} = \frac{1}{\sqrt{\tilde{F}}}  \left( \begin{array}{cc}
     \tilde{Q}_{2} & 0 \\
      0 & \tilde{Q}_{2} \end{array} \right) 
\ \ \ ; \ \ \
 \Vcal_{4} = \frac{1}{\sqrt{\tilde{F}}}  \left( \begin{array}{cc}
     \tilde{P}_{2} & 0 \\
      0 & -\tilde{P}_{2} \end{array} \right) 
$$
\be
\Vcal_{i>3}=0.
\label{e96}
\ee
The property $\Tr \left[ \Vcal_{i},\Vcal_{j}\right]=0$ implies the absence
of D2-branes. The configuration represents two D4-branes bound to a total
density $\sigma_{0}=2(\tilde{F}/2\pi)^{2}$ of D0-branes.

\subsection{D0 probe}
Let us first take the probe to be a D0-brane. This case has been considered
before in \cite{CT1} using different techniques, but we present it here in
order to illustrate our methods in a simple context.

 From supergravity, 
\begin{eqnarray}
S_{0} &=& -N_{1}T_{0}\int \! dt \ 
\{e^{-\phi}\sqrt{g_{MN}\dot{X}^{N}\dot{X}^{M}}-
C^{(1)}_{t}\} \nonumber \\
\mbox{} &=& -N_{1}T_{0}\int \! dt \ h^{-1}\{\sqrt{1-hv_{\parallel}^{2}-
hH_{4}v_{\perp}^{2}}-1 \} \label{e97} \\
\mbox{} &=& {\cal O}(h_{4}^{\,0})-
\frac{N_{1}N_{4}}{8}[\tilde{F}^{2}(v_{\parallel}^{2}+v_{\perp}^{2})^2
+4v_{\perp}^{2}]h_{4}+ {\cal O}(h_{4}^{\,2}). \nonumber
\end{eqnarray}
Here $v_{\parallel}^{2}=v_{1}^{\,2}+\cdots +v_{4}^{\,2} $ , 
 $v_{\perp}^{2}=v_{5}^{\,2}+\cdots +v_{9}^{\,2} $, and we keep only the one
loop contribution.

\mbox{} From Matrix theory
\be
\Ucal_{i} = (b_{i}+v_{i}t)\otimes \unitm_{2\times 2}
\ee
so,
\be
\frac{1}{2}{\cal F}_{MN}dX^{M}\wedge dX^{N} = 
(v_{i} \unitm_{N1\times N1} \otimes \unitm_{2\times 2})(dX^{0}\wedge dX^{i})
+ (\frac{1}{\tilde{F}} \unitm_{N1\times N1} \otimes \sigma_{3})
(dX^{1}\wedge dX^{2}+dX^{3}\wedge dX^{4}),
\ee
and 
\be
\hsf=Q_{0}\left[ \Ucal_{i}\otimes  \unitm_{2\times 2} -
  \unitm_{N_{1}\times N_{1}}\otimes \Vcal_{i}^{*}\right]^{-7}
= Q_{0} \, \unitm_{N_{1}\times N_{1}} \otimes 
\left( \begin{array}{cc} \alpha & 0 \\ 0 & \beta \end{array} \right)^{-7/2}
\ee
where
\begin{eqnarray}
\alpha &=& (b_{1}+v_{1}t-\frac{\tilde{Q}_{1}}{\sqrt{\tilde{F}}})^{2}+
 (b_{2}+v_{2}t-\frac{\tilde{P}_{1}}{\sqrt{\tilde{F}}})^{2}+
 (b_{3}+v_{3}t-\frac{\tilde{Q}_{2}}{\sqrt{\tilde{F}}})^{2}+
 (b_{4}+v_{4}t-\frac{\tilde{P}_{2}}{\sqrt{\tilde{F}}})^{2} \nonumber \\
 \mbox{} & \mbox{}& +(b_{\perp}+v_{\perp}t)^{2}
 \nonumber \\
\beta &=&  (b_{1}+v_{1}t-\frac{\tilde{Q}_{1}}{\sqrt{\tilde{F}}})^{2}+
(b_{2}+v_{2}t+\frac{\tilde{P}_{1}}{\sqrt{\tilde{F}}})^{2}+
 (b_{3}+v_{3}t-\frac{\tilde{Q}_{2}}{\sqrt{\tilde{F}}})^{2}+
 (b_{4}+v_{4}t+\frac{\tilde{P}_{2}}{\sqrt{\tilde{F}}})^{2} \nonumber \\
 \mbox{} & \mbox{}& +(b_{\perp}+v_{\perp}t)^{2}.
\end{eqnarray}
The Matrix theory action is
$$
S_{\rm{MT}}=
 {\cal O}(\hsf^0) -\frac{1}{8}T_{0}\,\Tr^{(N_{1})}\Tr^{(N_{2})}   
\int \! dt \ \hsf \left[ \tr (\eta \Fcal )^4 
  - \frac{1}{4}(\tr (\eta \Fcal )^2 )^2 \right] + {\cal O} (\hsf^2) \ .
$$
Plugging in the background we find,
$$
 \tr (\eta \Fcal )^4 - \frac{1}{4}(\tr (\eta \Fcal )^2 )^2 = 
\left[ (v_{\parallel}^{2}+v_{\perp}^{2})^2+
\frac{4 v_{\perp}^{2}}{\tilde{F}^{2}}\right]
\unitm_{N_{1}\times N_{1}}\otimes \unitm_{2 \times 2}.
$$
Now,
$$
\Tr^{(N_{1})} \rightarrow N_{1} \ \ \ ; \ \ \
\Tr^{(N_{2})} \rightarrow \frac{1}{(2\pi)^2}\int \!
 d\tilde{P}_{1}\, d\tilde{Q}_{1}\, d\tilde{P}_{2}\, d\tilde{Q}_{2}
\ \Tr^{(2\times 2)}.
$$
Defining
$$
\left( \tilde{\xi}^{1}, \tilde{\xi}^{2}, \tilde{\xi}^{3}, 
\tilde{\xi}^{4}\right) = 
\left(\frac{\tilde{P}_{1}}{\sqrt{\tilde{F}}},
\frac{\tilde{Q}_{1}}{\sqrt{\tilde{F}}},
\frac{\tilde{P}_{2}}{\sqrt{\tilde{F}}},
\frac{\tilde{Q}_{2}}{\sqrt{\tilde{F}}}\right)
$$
we have
$$
\Tr^{(N_{2})} \rightarrow \frac{\tilde{F}^2}{(2\pi)^2}\int \! d^{4}\tilde{\xi}
\ \Tr^{(2\times 2)}.
$$
We also need
\begin{eqnarray}
\frac{1}{(2\pi)^2} \int \! d^{4}\tilde{\xi} \ \hsf &=& \frac{Q_{0}}{(2\pi)^2}
\int \! d^{4}\tilde{\xi} \ \left[(\tilde{\xi}^{1})^2+
\cdots+(\tilde{\xi}^{4})^2
+(b_{\perp}+v_{\perp}t)^{2}\right]^{-7/2} \ \unitm_{2\times 2} \\
\mbox{} &=& 
\frac{Q_{0}}{15 (b_{\perp}+v_{\perp}t)^{3}} \ \unitm_{2\times 2} 
=\frac{Q_{4}}{r_{\perp}^{3}} \ \unitm_{2\times 2} =h_{4} \ \unitm_{2\times 2}.
\end{eqnarray}
Putting it all together we find
\be
S_{\rm{MT}}= {\cal O}(\hsf^{0})-\frac{2 N_{1}}{8}
[\tilde{F}^{2}(v_{\parallel}^{2}+v_{\perp}^{2})^2+4 v_{\perp}^{2}]h_{4}
+  {\cal O}(\hsf^{2})
\ee
in agreement with (\ref{e97}) when we set $N_{4}=2$.

\subsection{D(4+0) probe}
For our final example we will consider the interaction of branes oriented
at a relative angle \cite{BDL}.  The source 
will again be the (4+0) configuration
lying in the $(1234)$ plane.  For the probe, we start with a (4+2+2+0)
configuration in the $(1234)$ plane and then rotate in the $(15)$ plane by an
angle $\theta$.\footnote{A different configuration of D4-branes
at angles was recently studied in \cite{OZ}.}  
We will choose the  (4+2+2+0) state to be at the self-dual
point, so that the force between the branes will vanish when $\theta=0$.

We begin with supergravity. The  (4+2+2+0) probe is described by
$$
X^{0,1,2,3,4}= \xi^{0,1,2,3,4} \ \ ; \ \ X^{5}=\xi^{1}\tan\theta \ \ ; \ \
X^{6}=b_{6}
$$
$$
\frac{1}{2} F_{\mu\nu} d\xi^{\mu} \wedge d\xi^{\nu} = 
F_{12} d\xi^{1} \wedge d\xi^{2}+F_{34} d\xi^{3} \wedge d\xi^{4}.
$$
The probe action is
\begin{eqnarray}
S_{4}&=&-\frac{T_{0}}{(2\pi)^2}\int \! d^{5}\xi \  \left\{e^{-\phi}
\sqrt{\rm{det}[g_{MN}\partial_{\mu}X^{M}\partial_{\nu}X^{N}-F_{\mu\nu}]}
-F_{12}^{\,2}C_{t}^{(1)}-C_{t1234}^{(5)}\right\} \nonumber \\
\mbox{}&=& -\frac{T_{0}}{(2\pi)^2}\int \! d^{5}\xi \left\{ h^{-1}
[\sqrt{(hH_{4}^{-1}+F_{12}^{\,2})(hH_{4}^{-1}+F_{12}^{\,2}+h\tan\theta)}
-F_{12}^{\,2}]-H_{4}^{-1}+1\right\} \nonumber \\
\mbox{} &=& {\cal O}(h^{\,0}_{4})-\frac{T_{0}}{4(2\pi)^2}\int \! d^{5}\xi \ 
\frac{\tilde{F}^{2}}{F_{12}^{\,2}}\tan^{4}\!\theta \, h_{4} +  
{\cal O}(h_{4}^{\,2}) \label{e105}
\end{eqnarray}
where
$$
h_{4}= \frac{Q_{0}}{15 [b_{6}^{\,2}+(\xi^{1})^2\tan^{2}\!\theta]^{3/2}}.
$$

Now we consider Matrix theory.  The (4+2+2+0) probe background is
\begin{eqnarray}
\Ucal_{1} &=& \frac{Q_{1}}{\sqrt{F_{12}\cos\theta}} \ \ \ \ \  ; \ \ \ \ \
\Ucal_{2} = \frac{\sqrt{\cos\theta}P_{1}}{\sqrt{F_{12}}} \nonumber \\
\Ucal_{3}&=& \frac{Q_{2}}{\sqrt{F_{12}}} \ \ \ \ \ \ \ \ \ \ \ \ ; \ \ \ \ \  
\Ucal_{4} = \frac{P_{2}}{\sqrt{F_{12}}}  \\
\Ucal_{5}&=& \frac{\tan\theta P_{1}}{\sqrt{F_{12}\cos\theta}}  \ \ \ \ \ \ 
; \ \ \ \ \  
\Ucal_{6} = b_{6}, \nonumber
\end{eqnarray}
where we have included the $\theta$ dependence following the method described
in section 4.
Using the $\Vcal_{i}$ as in (\ref{e96}) we compute ${\cal F}_{MN}$:
\be
\frac{1}{2}{\cal F}_{MN}dX^{M}\wedge dX^{N} = 
(\frac{1}{F_{12}}\unitm_{2\times 2}- \frac{1}{\tilde{F}}\sigma_{3})
(dX^{1}\wedge dX^{2}+dX^{3}\wedge dX^{4})+ \frac{\tan\theta}{F_{12}}
\unitm_{2\times 2}dX^{1}\wedge dX^{5},
\ee
and 
\be
\hsf=Q_{0}[\Ucal_{i}\otimes \unitm_{2\times 2}-\Vcal_{i}^{*}]^{-7}=
Q_{0}\left( \begin{array}{cc} \alpha & 0 \\ 0 & \beta \end{array}
\right)^{-7/2}
\ee
with
\begin{eqnarray}
\alpha &=& \left(\frac{Q_{1}}{\sqrt{F_{12}}}-
\frac{\tilde{Q}_{1}}{\sqrt{\tilde{F}}}\right)^{2}+
 \left(\frac{P_{1}}{\sqrt{F_{12}}}-
\frac{\tilde{P}_{1}}{\sqrt{\tilde{F}}}\right)^{2}+
 \left(\frac{Q_{2}}{\sqrt{F_{12}}}-
\frac{\tilde{Q}_{2}}{\sqrt{\tilde{F}}}\right)^{2}+
 \left(\frac{P_{2}}{\sqrt{F_{12}}}-
\frac{\tilde{P}_{2}}{\sqrt{\tilde{F}}}\right)^{2} +b_{6}^{\,2}
 \nonumber \\
\beta &=& 
 \left(\frac{Q_{1}}{\sqrt{F_{12}}}-
\frac{\tilde{Q}_{1}}{\sqrt{\tilde{F}}}\right)^{2}+
 \left(\frac{P_{1}}{\sqrt{F_{12}}}+
\frac{\tilde{P}_{1}}{\sqrt{\tilde{F}}}\right)^{2}+
 \left(\frac{Q_{2}}{\sqrt{F_{12}}}-
\frac{\tilde{Q}_{2}}{\sqrt{\tilde{F}}}\right)^{2}+
 \left(\frac{P_{2}}{\sqrt{F_{12}}}+
\frac{\tilde{P}_{2}}{\sqrt{\tilde{F}}}\right)^{2} +b_{6}^{\,2}.\nonumber
\end{eqnarray}
We now find
\be
 \tr (\eta \Fcal )^4 - \frac{1}{4}(\tr (\eta \Fcal )^2 )^2 =
\frac{\tan^{4}\!\theta}{F_{12}^{\,4}}\unitm_{2\times 2}.
\ee
In a similar manner to the D0-brane probe example, we have
\begin{eqnarray}
\Tr^{(N_{1})} &\rightarrow& \frac{1}{(2\pi)^2}\int \!
 dP_{1}\, dQ_{1}\, dP_{2}\, dQ_{2}
\ \Tr^{(2\times 2)} =  \frac{F_{12}^2}{(2\pi)^2}\int \! d^{4}\xi
\ \Tr^{(2\times 2)} \nonumber \\
\Tr^{(N_{2})} &\rightarrow& \frac{1}{(2\pi)^2}\int \!
 d\tilde{P}_{1}\, d\tilde{Q}_{1}\, d\tilde{P}_{2}\, d\tilde{Q}_{2}
\ \Tr^{(2\times 2)} =  \frac{\tilde{F}^2}{(2\pi)^2}\int \! d^{4}\tilde{\xi}
\ \Tr^{(2\times 2)} 
\end{eqnarray}
and
\be
\frac{1}{(2\pi)^2}\int \! d^{4}\tilde{\xi} \ \hsf = 
\frac{Q_{0}}{15 [b_{6}^{\,2}+(\xi^{1})^{2}\tan^{2}\!\theta]^{3/2}}
\unitm_{2\times 2}
=h_{4}\unitm_{2\times 2}.
\ee
Plugging into $S_{\rm{MT}}$ we find,
\be
S_{\rm{MT}}= {\cal O}(\hsf^{0})-\frac{T_{0}}{4(2\pi)^2}\int \! d^{5}\xi
\frac{\tilde{F}^2}{F_{12}^{\,2}}\tan^{4}\!\theta\,h_{4}+ {\cal O}(\hsf^{2})
\ee
in agreement with the supergravity result (\ref{e105}).

\section{Conclusions}

In this  work we have shown that a large class of D-brane probe actions can
be recovered from a simple ansatz for the Matrix theory effective action. We
subjected our ansatz to a number of highly demanding consistency checks and
found the correct behavior in all cases.  In the case of D0-brane sources we
were able to perform the checks to all loop orders, whereas we were restricted
to one loop in the case of D4-brane sources.  The obstacle which prevented
us from extending the latter results beyond one loop was  the presence of 
ambiguities in evaluating the trace over the source variables.  The resolution
of the trace ambiguities would constitute a significant advance, and would
allow us to make contact with, for example, the discrepancies reported in
\cite{DouPolStr}.  One would also like to be able to treat fully non-abelian
backgrounds, instead of having to assume, as was the case here, that 
commutators of field strengths are small.

It is worth pointing out some differences between the approach to scattering
in Matrix Theory developed here and other approaches in the literature.  In
the majority of cases appearing elsewhere, one computes first the phase shift
of the probe rather than its effective Lagrangian.  Having a formalism in 
which the Lagrangian appears directly, as it does here, is a great advantage
when one treats complicated processes involving, for instance, non-extremal
branes.  Computing the phase shift in such circumstances would be prohibitively
difficult.  Another difference with other work concerns the use of T-duality.
Many treatments take the branes to be wrapped on a torus, perform T-duality,
and use the relation between Matrix Theory and 
super Yang-Mills field theory \cite{BFSS,WT}.
In contrast, we always work in the original spacetime.  The latter approach
seems to us to be simpler and to make the physical picture more readily
visualizable.

It would be interesting to generalize our methods to include non-extremal
sources, similar to what was done  in \cite{Mal}.  The novel feature in this 
case
is that one has to average over all degenerate backgrounds in order to compare
with supergravity results.  Perhaps Matrix theory can shed light on this 
intriguing situation. For instance, it is possible to include both non-extremal
probes and sources into the MT effective action (\ref{e17}) and do at least
one loop calculations.

Finally, it is suggestive that our ansatz is as simple as it is.  It might be
expected that, with a proper understanding, one could derive it directly from
first principles without having to explicitly evaluate an infinite series of
loop diagrams.  

\bigskip

\bigskip

\noindent
{\large {\bf Note Added}}

\bigskip

\noindent
As this work was being completed, there appeared ref. \cite{CT3} where 
similar ideas were developed independently.

\bigskip

\bigskip

\noindent
{\large {\bf Appendix: Null Reduction}}

\bigskip

\noindent
In this appendix we present, for convenience, the null reductions of the 
D0-brane and D(4+0)-brane metrics.  The method we use is to start with the
standard
ten dimensional backgrounds, lift them up to eleven dimensions using the
relation \cite{TfourM}
\be
ds_{11}^{2}=e^{-2\phi/3}ds_{10}^{2}-e^{4\phi/3}(dx^{11}-dx^{M}C_{M})^{2},
\ee
and then reduce back to ten dimensions along a null direction.  

\bigskip

\noindent
{\large {\bf A.1 D0-brane}}

\bigskip

\noindent
In ten dimensions
\begin{eqnarray}
ds_{10}^{2}&=&H_{0}^{-1/2}dt^2-H_{0}^{1/2}dx^{i}dx^{i} \nonumber \\
e^{-\phi}&=&H_{0}^{-3/4} \ \ \ ; \ \ \ C_{t}=H_{0}^{-1}-1.
\end{eqnarray}
Lifting to eleven dimensions we find
\be
ds_{11}^{2}=dt^2-(dx^{11})^{2}-h(dx^{11}-dt)^{2}-dx^{i}dx^{i}
\ee
where we have defined $h=H_{0}-1$.
Transforming to null coordinates $x^{\pm}=x^{11}\pm t$, and denoting 
$\tau=x^{+}/2$, the eleven dimensional metric appears as
\begin{eqnarray}
ds_{11}^{2}&=&2 d\tau dx^{-}-h dx^{-} dx^{-}-dx^{i} dx^{i} \nonumber \\
\mbox{}&=&e^{-2\phi/3}ds_{10}^{2}-e^{4\phi/3}(dx^{-}-C_{\tau}d\tau)^{2}.
\end{eqnarray}
\mbox{} From the above we deduce that 
$e^{-\phi}=h^{-3/4}$, $C_{\tau}=h^{-1}$.  The
null reduced metric in ten dimensions is then \cite{BBPT}
\be
ds_{10}^{2}=h^{-1/2}d\tau^{2}-h^{1/2}dx^{i}dx^{i}.
\ee
The difference between this metric and the starting metric is that the 
harmonic function $H_{0}$ has been transformed into $h$.

\bigskip

\noindent
{\large {\bf A.2 D(4+0)-brane}}

\bigskip

\noindent
The ten dimensional metric of the marginally bound (4+0) state \cite{ATM} is
\begin{eqnarray}
ds_{10}^{2}&=&H_{0}^{-1/2}H_{4}^{-1/2}dt^{2}
-H_{0}^{1/2}H_{4}^{-1/2}(dx_{1}^{2}+\cdots + dx_{4}^{2})
-H_{0}^{1/2}H_{4}^{1/2}(dx_{5}^{2}+\cdots + dx_{9}^{2}) \nonumber \\
e^{-\phi}&=&H_{0}^{-3/4}H_{4}^{1/4} \ \ \ ; \ \ \ C_{t}=H_{0}^{-1}-1.
\label{e118}
\end{eqnarray}
Lifting to eleven dimensions gives
\be
ds_{11}^{2}=H_{4}^{-1/3}\left[dt^{2}-(dx^{11})^{2}-h(dx^{11}-dt)^{2}\right]
-H_{4}^{-1/3}(dx_{1}^{2}+\cdots + dx_{4}^{2})
-H_{4}^{2/3}(dx_{5}^{2}+\cdots + dx_{9}^{2}).
\ee
Transforming to null coordinates and reducing to ten dimensions in the same
way as for the D0-brane case yields
\begin{eqnarray}
ds_{10}^{2}&=&h^{-1/2}H_{4}^{-1/2}d\tau^{2}
-h^{1/2}H_{4}^{-1/2}(dx_{1}^{2}+\cdots + dx_{4}^{2})
-h^{1/2}H_{4}^{1/2}(dx_{5}^{2}+\cdots + dx_{9}^{2}) \nonumber \\
e^{-\phi}&=&h^{-3/4}H_{4}^{1/4} \ \ \ ; \ \ \ C_{\tau}=h^{-1}.
\end{eqnarray}
Comparing with (\ref{e118}), we see that 
the harmonic function $H_{0}$ for the D0-brane 
has been replaced by $h$, while the harmonic function $H_{4}$ keeps its 
original form.

\end{document}